\documentstyle[prb,subeqn,aps,psfig]{revtex}

\begin{document}
\draft
\title{\bf Breathers in a single plaquette of Josephson junctions : \\
existence, stability and resonances.}
\author{ A. Benabdallah, M. V. Fistul and S. Flach}
\address{Max-Planck-Institut f\"ur Physik komplexer Systeme,
N\"othnitzer Stra\ss e 38, D-01187 Dresden, Germany}
\date{\today}
\wideabs{ 
\maketitle
\begin{abstract}
We present a theoretical study of {\it inhomogeneous} dynamic (resistive) 
states in a single plaquette consisting of three Josephson junctions.
These breather states are found to exist
in a large range of
control parameters (dc bias $\gamma$, anisotropy $\eta$ and self inductance 
$\beta_L$).
We perform an analytical analysis of these states, and their 
stability and resonance behaviour. Numerical calculations confirm
our results and show that instabilities arise due to
resonant interactions between the breather and 
electromagnetic oscillations (EOs) mediated by the breather state. 
These instabilities manifest themselves by various nonlinearities,
resonant steps and voltage jumps in the current-voltage characteristics. 
\end{abstract}
}
\section{Introduction}
In recent years we witnessed a considerable advancing
in the area of
study of localized excitations in nonlinear lattices.
\cite{Aubry97,fw98,SivPage} These excitations are coined
discrete breathers and are time-periodic and spatially
localized solutions of the underlying dynamical equations
of motion. 
They generically appear in spatially homogeneous systems 
due to the interplay between discreteness and nonlinearity. 
Discrete breathers were detected experimentally in different
systems, e.g. in weakly coupled optical waveguides, \cite{eisenberg} 
in the lattice dynamics of solids, \cite{Bishop} in
antiferromagnets \cite{Sievers} and in systems of interacting Josephson
junctions. \cite{tmo00,Binder00,bau00,tmbo00,marcus}

Especially the latter example served for many years as a well controlled
laboratory object to study various nonlinear phenomena.
Josephson junction systems allow for an adequate
theoretical description of the underlying dynamics. 
The originally proposed and realized ladder geometry of
small interacting Josephson junctions with an external dc bias
became a system of intense theoretical and experimental studies. 
\cite{Floria96,Spain98,Spicci99,mto99} 
The dynamic localized states in Josephson junction ladders persist 
due to a well known property of a single Josephson junction, namely
the presence of two different states.
These are a superconducting state (zero voltage drop across
the junction) or
a resistive state (nonzero voltage drop across the junction).
If the junction is underdamped, both states may coexist in 
a wide range of an externally applied dc bias.
A well known consequence of that 
is a hysteretic current-voltage ($I-V$) characteristics.

A breather state in such a ladder is characterized by a few junctions
being in the resistive state, while the rest of all junctions
reside in the superconducting one. These inhomogeneous dynamic 
states lead to specific total dc voltage drops across the ladder 
which are used to plot $I-V$ characteristics. This method
has been successfully
combined with snapshots made using the low temperature scanning laser
microscopy technique (LTSM) 
which allows for a spatial resolution of dc voltage drops. \cite{Binder00,bau00} 
Note that the experimentally accessible information
provides only with time averaged 
data, so the internal
dynamics of breather states is so far not directly observable in 
experiments.
The full dynamical picture
is however more subtle. E.g. the superconducting junctions display 
nonzero ac voltage drops. The amplitude of these ac voltages
decays to zero with increasing spatial distance from the resistive
junctions. Moreover, a breather state may be tuned into resonance
with other dynamical modes of the system which leads to an increasing
complexity of the breather dynamics.  
Experimental studies revealed a whole ``zoo'' of different 
breather states, instabilities,
resonances and switching scenarios depending on the parameters of the
ladder. \cite{tmo00,Binder00,bau00,tmbo00,marcus} Similar features have been found in numerical simulations.

To clarify the role of control parameters and the cause
of the primary (generic) instabilities, resonances and switchings
we 
carry out a consistent 
study of a {\it single plaquette model} containing three Josephson junctions.
Whereas the occurrence of breathers in an extended ladder may be 
interpreted as a dynamical breaking of discrete translational invariance,
in the single plaquette model the analogy of the translational
invariance becomes a permutational invariance. It is this permutational
invariance which will be violated by dynamical states coined breathers
in analogy to the ladder case. The approach of studying inhomogeneous 
dynamic states 
in reduced systems \cite{mto99,MFrowsw} has been successfully used in the past 
to get
deeper understanding for different breather properties as e.g. 
energy radiation and quantization. \cite{fw98} 

Such a single plaquette model is the simplest
system which supports permutationally broken 
states in the presence of a homogeneous dc bias $\gamma$ (Fig.1).
It consists of two {\it vertical} Josephson junctions parallel to
the bias direction and a {\it horizontal} one that is transversal to the bias.
The Josephson junctions are indicated by crosses.
In a time-averaged picture we expect only four different
states to be realized. The first one is characterized by
all junctions being in the superconducting state (SS), with zero
current flowing through the horizontal junction.
The second one is the homogeneous whirling state (HWS) when both vertical
junctions are in the resistive state. Due to 
permutational invariance of this state the current flowing 
through the superconducting
horizontal junction is still zero. 
Note that both states are invariant under permutation, i.e. the exchange of two 
vertical junctions.
The states of interest are the remaining
two breather states. Each of these states is characterized by only
one vertical junction and in addition the horizontal junction being resistive.
The second vertical junction is in a superconducting state, although there is
a nonzero current flowing through it. Obviously both breather states
are not invariant under permutation.

In the following we will systematically 
study the dependence of breather state properties
on the control parameters of the system (dc bias, anisotropy, self-inductance). 
Our paper is organized in the following way. 
In section II we present the model, introduce the equations
of motion and provide with an approximative analytical
treatment. 
In section III we introduce a number of
numerical methods for computing breathers, obtaining their stability and
following their evolution. With their help we find the domains
of existence, stability and resonances of breather states
in the space of control parameters and relate them to computed
$I-V$ characteristics.

\begin{figure}[!hbp]
\vskip 0.6cm

\centerline{\psfig{figure=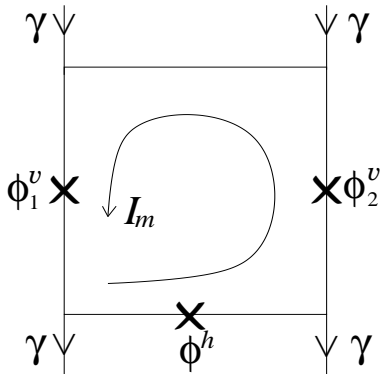,angle=-90,width=5cm,height=5cm}}
\caption{Sketch of the plaquette with three Josephson junctions 
(marked by crosses) in the presence
of uniform dc bias. Arrows indicate the direction of external current flow 
(dc bias $\gamma$).}
\end{figure}

\section{Equations of motion and properties of breather solutions}

\subsection{Equations of motion}

The dynamics of a single plaquette of three Josephson junctions 
(see Fig. 1) is determined by the time dependent Josephson phases of 
vertical junctions $\phi^{v}_{1,2}$, and the horizontal junction 
$\phi^{h}$. We will consider an anisotropic system and the parameter  
of anisotropy is $\eta~=~ I_{ch}/I_{cv}$, where 
 $I_{ch}$ and $I_{cv}$ are respectively the
critical currents of the horizontal and vertical junctions.
By making use of the Resistive Shunted Junction model for each junction 
\cite{Likharev86} we obtain the following set of equations
\begin{eqnarray}
\nonumber
{\cal N} (\phi^{v}_1) &=& \gamma + I_{ m} \; , \\ \label{phi1}
{\cal N} (\phi^{v}_2) &=& \gamma - I_{ m} \; , \\ \nonumber
{\cal N} (\phi^{h}) &=& I_{ m}/\eta \; ,
\end{eqnarray}

The nonlinear operator ${\cal N}$ is defined as 
\begin{equation}
{\cal N}(\phi) ~=~ \ddot \phi +
\alpha \dot \phi + \sin (\phi) \; .
\label{2-2} 
\end{equation}
Here, the unit of time is the inverse plasma frequency,
the dc bias $\gamma$ is normalized to the critical current of 
the vertical junctions
$I_{cv}$, and 
the dimensionless parameter $\alpha$ determines the damping in the system.
In (\ref{phi1}) we introduce 
the mesh current $I_{ m}$ that in turn, is determined by the flux 
quantization law \cite{Barone82}
\begin{equation}
\label{self}
-\beta_L I_{ m}~=~\phi^{v}_1 - \phi^{v}_2 + \phi^{h} \; ,
\end{equation}
where $\beta_L ~=~2\pi I_{cv} L/\Phi_0$ is the normalized 
self-inductance $L$ of the plaquette.  
Note here, that the voltage drop across a junction is given by $\dot{\phi}$. 

\subsection{Breather solutions}

We are interested in the properties of 
inhomogeneous breather states.  
In the limit $\eta=0$ we may exactly solve the equations of motion
and implicitly obtain the breather solution in terms of
the resistive state of a single vertical junction. In this case
the mesh current vanishes $I_m=0$ and the two vertical junctions
decouple, while the horizontal one follows the dynamics of
the vertical junctions (below we will consider a particular breather case as 
the left vertical junction is in the superconducting state):
\begin{eqnarray}
\phi^{v}_1 &=& {\rm arcsin}\; \gamma~~,
\nonumber \\
\phi^{v}_2 &=& f(\alpha,\gamma,t)~~,
\label{2-11} \\
\phi^{h} &=& f(\alpha,\gamma,t) - {\rm arcsin}\; \gamma~~.
\nonumber
\end{eqnarray}
Here, the function $f(\alpha,\gamma,t)$ describes the evolution
of a single vertical junction in the resistive state with bias $\gamma$
and damping $\alpha$. It is well known that this state exists as 
\cite{Likharev86}
\begin{equation}
\gamma \ge \gamma_r=\frac{4 \alpha}{\pi}~~,
\label{2-ret}
\end{equation}
where $\gamma_r$ is the retrapping current of a single Josephson junction. 
On the other side
the superconducting state of the first vertical junction 
(see Eq. (\ref{2-11})) will 
persist only if $\gamma < 1$. Consequently we obtain that the breather
solution (\ref{2-11}) exists for $\eta=0$ provided $\gamma_r \le \gamma \le 1$.

For nonzero values of $\eta \neq 0$  exact analytical solutions are 
not available. In the following we obtain approximate solutions using a  
{\it dc analysis}, i. e. neglecting the Josephson phase oscillations. 
We write the Josephson phases in the form:
\begin{eqnarray}
\nonumber
\phi^{v}_1 &\approx&  c^{v}_1~~, \\ \label{phiapprox}
\phi^{v}_2 &\approx& \Omega t + c^{v}_2~~, \\ \nonumber
\phi^{h} &\approx& \Omega t + c^{h} \, ,
\end{eqnarray}
where $\Omega$ is the {\it breather frequency}. Substituting these  
expressions into (\ref{phi1}) and (\ref{self}) the mesh current reads
\begin{equation}
I_{ m}  = \frac{\gamma \eta}{1+\eta}~~. 
\label{2-8}
\end{equation}
Correspondingly the dc bias dependent breather frequency and the Josephson 
phase shifts are
\begin{eqnarray}
\label{cur_fre}
\Omega &=&  \frac{\gamma}{\alpha(1+\eta)}~~, \\ \label{cun}
c^{v}_1 &=& {\rm asin} \left ( \frac{1+2\eta}{1+\eta} \gamma \right )~~, \\ 
\label{2-9c}
c^{v}_2 &=& c^v_1 +
\frac{\beta_L \gamma \eta}{1+\eta}~~.
\end{eqnarray}
Here, we may choose
$c^{h}=0$ without loss of generality.
As a result we find the
$I-V$ characteristics of the Josephson junction plaquette in the 
presence of a breather state:
\begin{equation}
\gamma~=~V \alpha (1+\eta)~~,
\label{IV}
\end{equation}
where $V~=~<\dot \phi^{v}_2>~=~ \lim_{t \to \infty} \frac 1t 
\left(\phi^{v}_2(t)-\phi^{v}_2(0)\right)$.
The $I-V$ characteristics of a breather (\ref{IV}) differs
from that of the HWS which is  $\gamma~=~V \alpha$.
The breather is characterized by a smaller voltage drop across
the resistive vertical junction for a given current value $\gamma$.

As the dc bias current decreases the breather state can disappear as a 
solution of the dynamic equations and the system undergoes a transition to 
the superconducting state. By making use of the dc analysis and a standard 
theory of the retrapping current in a single small Josephson junction 
\cite{Likharev86} we expect that the breather state disappears as the current 
flowing through the horizontal (resistive) junction is equal $\gamma_r$.
This happens for the following value of the external dc bias $\gamma_{b,r}$: 
\begin{equation}
\label{ir}
\gamma_{b,r} = \alpha(1+\eta)\frac{4}{\pi}~~.
\end{equation}
Remarkably this value of the retrapping current does not
depend on the self inductance parameter $\beta_L$. The dependence
of $\gamma_{b,r}$ on $\eta$ is shown in Fig.2. 

For large values of the dc bias the breather state {\it switches} to the HWS.  
This switching occurs because the left vertical junction can not 
anymore support the 
superconducting state. From (\ref{cun}) we obtain that this switch should
take place at
\begin{equation}
\label{hws}
 \gamma = \frac {1+\eta}{1+2\eta}~~.
\end{equation}

\begin{figure}[!hbp]
\centerline{\psfig{figure=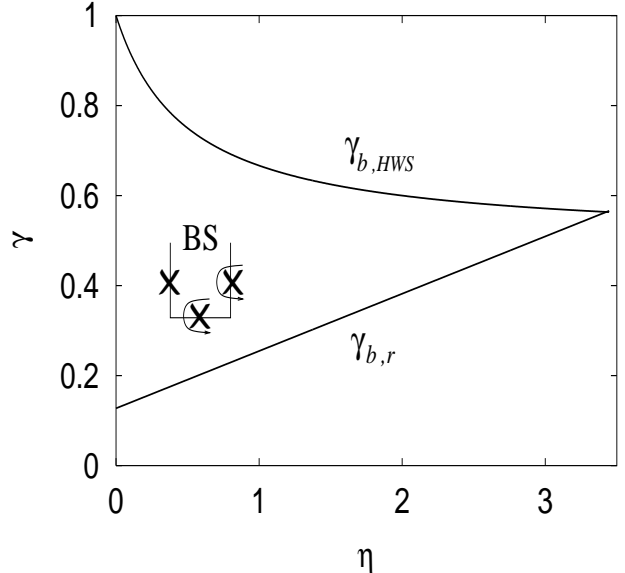,angle=-90,width=9cm,height=8cm}}
\caption{Analytic boundaries of the  breathers existence in $(\gamma,\eta)$ 
plane
for $\beta_L=0,$ $\alpha=0.1$. The dc analysis was used.}
\end{figure}

The crucial assumption of the above carried out analysis is 
the neglecting of the Josephson phase oscillations. Indeed, the inhomogeneous 
dynamic breather state induces the electromagnetic oscillations (EOs) in the 
plaquette and 
in turn, these EOs change the simple breather solution (\ref{phiapprox}). 
However, if the breather frequency $\Omega$ is large with respect to the 
characteristic frequencies of EOs, the influence of such a {\it nonresonant} 
interaction between the breather and EOs is  small. Thus, in the 
nonresonant case the amplitude $A$ of the breather induced EOs is
$ A~\simeq~\frac{\sqrt{2} \eta}{(2\eta+1) \Omega^2}$ and correspondingly, 
it leads to a corrected prediction of the 
critical dc bias value $\gamma_{b,HWS}$ where the breather switches
to the HWS:
\begin{equation} \label{NonresonantSwitch}
\gamma_{b,HWS}~=~\frac{1+\eta}{1+2\eta}(1-
\frac{\eta^2 (2\eta+1)^2 \alpha^4}{2})~~.
\end{equation}
The result is shown in Fig.2. Note that similar
existence regions for the breathers on extended ladders have been
reported in. \cite{tmbo00}

The main outcome of the presented analytical treatment is a triangular shaped 
region
in the space of control parameters $\gamma$ and $\eta$ where
we expect to find breather states. Notably the prediction is that
for small values of damping ($\alpha \ll 1$) the breather should exist
for rather large values of the anisotropy $\eta$, e.g. for
$\alpha=0.05$ the anisotropy can be as large as $\eta \sim 5$.
Such parameter regions are the opposite of what has been initially
anticipated, when the breather states were found by making use of the 
continuation arguments for the solution
(\ref{2-11}) at $\eta=0$ to small nonzero $\eta$ values. \cite{ma94} In this 
case 
the horizontal junctions serve as weak links. Instead we 
predict the existence of breather states for cases of large $\eta$
values where the vertical junctions serve as weak links, and the horizontal
one is a strong link!

\subsection{EO frequencies and resonances}

Next we turn to the analysis of the {\it resonant interaction} 
between the breather state and small amplitude EOs. 
In order to obtain the characteristic  
frequencies of EOs we linearize the dynamic equations 
around the breather state (\ref{phiapprox}):
 \begin{eqnarray}
  \label{beta1}
  \ddot \delta^{v}_1 + \alpha \dot \delta^{v}_1 + 
  \cos(c^{v}_1) \delta^{v}_1 
  &=&
  -\frac {1}{\beta_L} (\delta^{v}_1 - \delta^{v}_2 + \delta^{h})~~, 
  \\ \label{beta2}
  \ddot \delta^{v}_2+ \alpha \dot \delta^{v}_2 
  + \cos(\Omega t + c^{v}_2) \delta^{v}_2&=&
  \frac {1}{\beta_L} (\delta^{v}_1- \delta^{v}_2+ \delta^{h})~~, 
  \\ \label{beta3}
  \ddot \delta^{h} + \alpha \dot \delta^{h} 
  + \cos(\Omega t) \delta^{h}
  &=&
  -\frac {1}{\eta \beta_L} (\delta^{v}_1- \delta^{v}_2+ 
  \delta^{h})~~.
  \end{eqnarray}
  Here $\delta^v_{1,2}(t),\, \delta^{h}(t)$ correspond to small amplitude EOs
  excited in the presence of the breather state. 
  In the weakly damped case ($\alpha~\ll~1$) we may 
  neglect the damping term and 
  skip the time-periodic terms in
  (\ref{beta2}) and (\ref{beta3}). As a result we derive 
  two characteristic frequencies of EOs:
\begin{eqnarray}
\label{omega34}
|\omega_{\pm}| = \sqrt{ F \pm \sqrt{ F^2-G}} \;\;, \\
F~=~\frac{1}{2}\cos(c^{v}_1) + \frac {1 + 2\eta}{2\eta\beta_L}
\;\;, \label{3-F} \\
G~=\cos(c^{v}_1) \frac{(1 + \eta) }{\eta\beta_L}
\;\;. \label{3-G}
\end{eqnarray}
These frequencies depend on the anisotropy $\eta$ and the bias 
current through the phase shift $c_1^v$ (Eq.(\ref{cun})).
For large values of $\beta_L$ the lower frequency  
$\omega_{-} \sim \frac{1}{\sqrt{\beta_L}}$ 
and the large frequency $\omega_{+} \sim \sqrt{\cos(c^{v}_1)}$. 
In the opposite limit of small $\beta_L$ we find
$\omega_{+} \sim \frac{1}{\sqrt{\beta_L}}$ while 
$\omega_{-}~\sim
((\frac{1+\eta}{1+2\eta})^2-\gamma^2)^{1/4}$.

Since the breather is a time-periodic solution and the EOs
describe small amplitude deviations from the breather state,
we can expect that the following scenario holds. In the case of a
proper frequency matching the breather may drive EOs into
resonance. In turn the 
increase in the EO amplitudes
will proceed on the expense of the breather state. In other
words the possibility of a breather mediated resonance for EOs
turns into a resonant interaction between the breather and the EOs.
In such a case we expect in the full dynamical system 
various instabilities of the breather
state and corresponding switchings (the voltage jumps) 
or {\it resonant steps} in the $I-V$ characteristics
to occur.

Let us classify the possible expected resonance cases. 
The first class of {\it primary} resonances may appear
if 
\begin{equation}
\label{res1}
\omega_{\pm}  =   m \Omega
\end{equation}
where $m$ is an integer. These primary resonances correspond
to a matching of the breather frequency 
$\Omega$ or its multiples (higher harmonics
of the breather state)
with the EOs frequencies.

A second class consists of {\it parametric} resonances which 
are characterized by
\begin{equation}
\label{res2}
\omega_{\pm} =   (m + {1 \over 2}) \Omega \;\;.
\end{equation}

Finally we may also expect the appearance of {\it combination}
resonances occurring  as
\begin{equation}
\label{res3}
m \Omega = \omega_{-} \pm  \omega_{+}
\end{equation}
In such a case, which is at variance with the first two types of
resonances, the breather mediates a resonant interaction
between two different EO frequencies.

All listed scenarios of resonance involve the breather frequency
and the EO frequencies. While the approximate expression for
the breather frequency (\ref{cur_fre}) is {\it independent} on
$\beta_L$, the EO frequencies {\it depend} sensitively  on 
the self inductance (\ref{omega34}). Consequently the expected
positions of resonances vary with $\beta_L$. Since
the existence window for breathers (cf. Fig.2) is expected to not depend
on $\beta_L$,
certain
resonances may simply be shifted outside of the breather existence
region and thus, become irrelevant. 
E.g. for small values of $\beta_L$ the
EO frequency $\omega_{+}$ becomes very large and we expect then
only primary and parametric resonances involving $\omega_{-}$. 
If however intermediate values for $\beta_L$ are chosen,
additional primary and parametric resonances involving
$\omega_{+}$ appear. But most important we expect combination
resonances to appear, which can be realized only if the two
EO frequencies, their sum or difference are of the same order
as the breather frequency.

\section{Numerical analysis of breather dynamics}

\subsection{Description of numerical methods}

Let us describe the numerical methods of monitoring and
characterizing breather states. In all cases a Runge-Kutta
integration method of the differential equations (\ref{phi1})
is chosen. The first way is to fix all parameters including the
bias $\gamma$, to choose some proper initial conditions and to
integrate the equations for sufficiently long time until
the system relaxes onto a breather attractor. Once the attractor
is reached (with sufficient accuracy) all characteristics of this
state may be obtained, including the average voltage drop $V$. 
After that we may perform a small step in $\gamma$ and repeat the
procedure. Then we find the $I$-$V$ characteristics.
This method is especially good when approaching an instability,
since we obtain explicitly the switching to another state.
We also generate time-resolved images (movies) of the
full dynamical behaviour in order to visually check the realized
state or monitor the process of switching.
The drawback of this method is that it is not capable of explaining
the nature of an observed instability. In addition it may become
time consuming if the relaxation times become large.

Another way of studying the system states is to use a Newton scheme
that allows to iteratively find time-periodic states, e.g. breathers.
This method is very fast and possesses a high precision. 
Once a solution is found, it may be characterized similarly to
the way described above.
The 
drawback here is that we can not predict the outcome of a switching
if a breather state disappears or turns unstable. The method is
even insensitive to the stability property of a state under
consideration, so we may even miss the switching.

In order to obtain reliable information about the stability
properties of breather states we perform in addition
a Floquet analysis of small amplitude deviations from a breather
state which we generated with the help of a Newton scheme.
We consider the time evolution of small
perturbations  $\{\epsilon_i(0),\dot \epsilon_i(0)\}$ ($i=1,2,3$)
at $t=0$
of a breather solution $(\phi_{1,2}^{v}, \phi^{h})$
over one period $T_b$ of the breather state.
The outcome is a map of the space of small perturbations onto
itself, which is characterized by a 
Floquet matrix ${\cal M}$ (of rank 6):
\begin{equation}
\nonumber
\left\{
\begin{array}{cc}
\epsilon_i(T_b) \\
\dot \epsilon_i(T_b)
\end{array} \right \}
= {\cal M}
\left\{
\begin{array}{cc}
\epsilon_i(0) \\
\dot \epsilon_i(0)
\end{array} \right \}  \;\;. 
\end{equation}
This matrix is generated numerically.
The breather will be linearly stable if and only if
the matrix ${\cal M}$ has no eigenvalues with modulus larger
than one.
The quasi-symplectic properties of this map impose 
the following conditions on its eigenvalues:
if $\lambda$ is an 
eigenvalue, then $\lambda^*$, ${\rm e}^{-\alpha T_b}/\lambda$  
and ${\rm e}^{-\alpha T_b}/\lambda^*$ are also eigenvalues
(here $a^*$ denotes complex conjugation of $a$). \cite{Arnold}
Furthermore there is always one eigenvalue $\lambda_t=1$
which corresponds to perturbations tangent to the trajectory of the breather 
solution.
Consequently there always exists a second eigenvalue 
$\tilde{\lambda}_t = {\rm e}^{-\alpha T_b}$. These two
eigenvalues will not be of further interest.
The remaining four eigenvalues can be grouped into two pairs
each consisting of an eigenvalue and its complex conjugated partner.
These two groups correspond to the two EO frequencies (\ref{omega34}).
In the normal case of a stable breather two eigenvalues are simply
characterized by
\begin{equation}
\lambda^{\pm}={\rm e}^{-\alpha T_b/2} {\rm e}^{i \omega_{\pm} T_b}
\;\;.
\label{lambdapm}
\end{equation}
Thus the four relevant eigenvalues will be located on an inner circle
with radius $ {\rm e}^{-\alpha T_b/2} < 1$ in the complex plane.
The variation of parameters system leads to a change (moving) of 
Floquet eigenvalues.
Resonances correspond to collisions of an eigenvalue with
its complex conjugated partner on the positive real axis (primary resonance),
on the negative real axis (parametric resonance) 
or to the collision of two different
eigenvalues away from the real axis (combination resonance).
In all cases colliding eigenvalues will generally leave the inner circle,
one moving inside and one outside. Upon further variation of some
parameter the outer one may cross the unit circle, which marks the
point of instability of the breather state. For small values of
$\alpha$ the two events - collision and crossing the unit circle -
take place close to each other, so that a clear connection between
instabilities and resonances may be drawn.
The Floquet analysis thus allows in addition to characterize
the type of observed instability (resonance).

A combination of all three methods - direct numerical integration, 
Newton scheme and Floquet analysis - turns out to be a powerful tool which
provides with good insight of the dynamical processes under consideration.

\subsection{Domains of breather existence and instability windows.}

In order to obtain the breather solution for different values of anisotropy 
$\eta$ and correspondingly, the 
domains of breather existence in the parameter space $(\gamma,\eta)$, 
we used the following procedure. First, we 
fixed the value of the external dc bias 
$\gamma$ and found a breather solution in the form 
(\ref{2-11}). After that we increased the parameter $\eta$ by a small step
$\Delta \eta =10^{-2}$, and the previously found solution was used 
as an initial 
guess for a Newton method. The procedure was repeated for other values of 
dc bias $\gamma$. We varied the self inductance parameter 
$\beta_L$ while in most of our simulations the damping parameter 
$\alpha=0.1$ was fixed. 

\subsubsection{ Zero self inductance case ( $\beta_L=0).$}

In the particular case of zero self inductance $\beta_L$ the equations 
(\ref{phi1}) are simplified as $\phi^h~=~\phi_1^v-\phi_2^v$, and the system 
is characterized by just two degrees of freedom. 
We found that the domains of 
breather existence in the parameter space $\gamma$-$\eta$ have 
a triangular form, 
and it is shown in Fig. 3. 

One can also see that 
the existence domain expands in the region of large $\eta$ as the damping 
parameter $\alpha$ decreases (Fig. 3a and Fig. 3b). 
The maximum value of the anisotropy $\eta_{max}$ still allowing to obtain  
the breather state is estimated as $\displaystyle{\eta_{max}~\simeq~\frac{1}{4\alpha}}$
(see inset in Fig. 3b).
Note here, as the anisotropy is not large 
($\eta~\leq~1.5$ for the system with damping $\alpha~=~0.1$) 
the dc analysis developed in the Section II (see, Fig. 2 and 
Eqs. (\ref{ir}), (\ref{hws}) ) gives a good approximation for numerically 
simulated domain of breather existence (Fig. 3). 
As the anisotropy increases, $\eta~>~1$, the region of dc bias $\gamma$ 
where the breather state was found, shrinks with respect to the one 
obtained by making use of the dc analysis.
 
\begin{figure}[!hbp]
\centerline{\psfig{figure=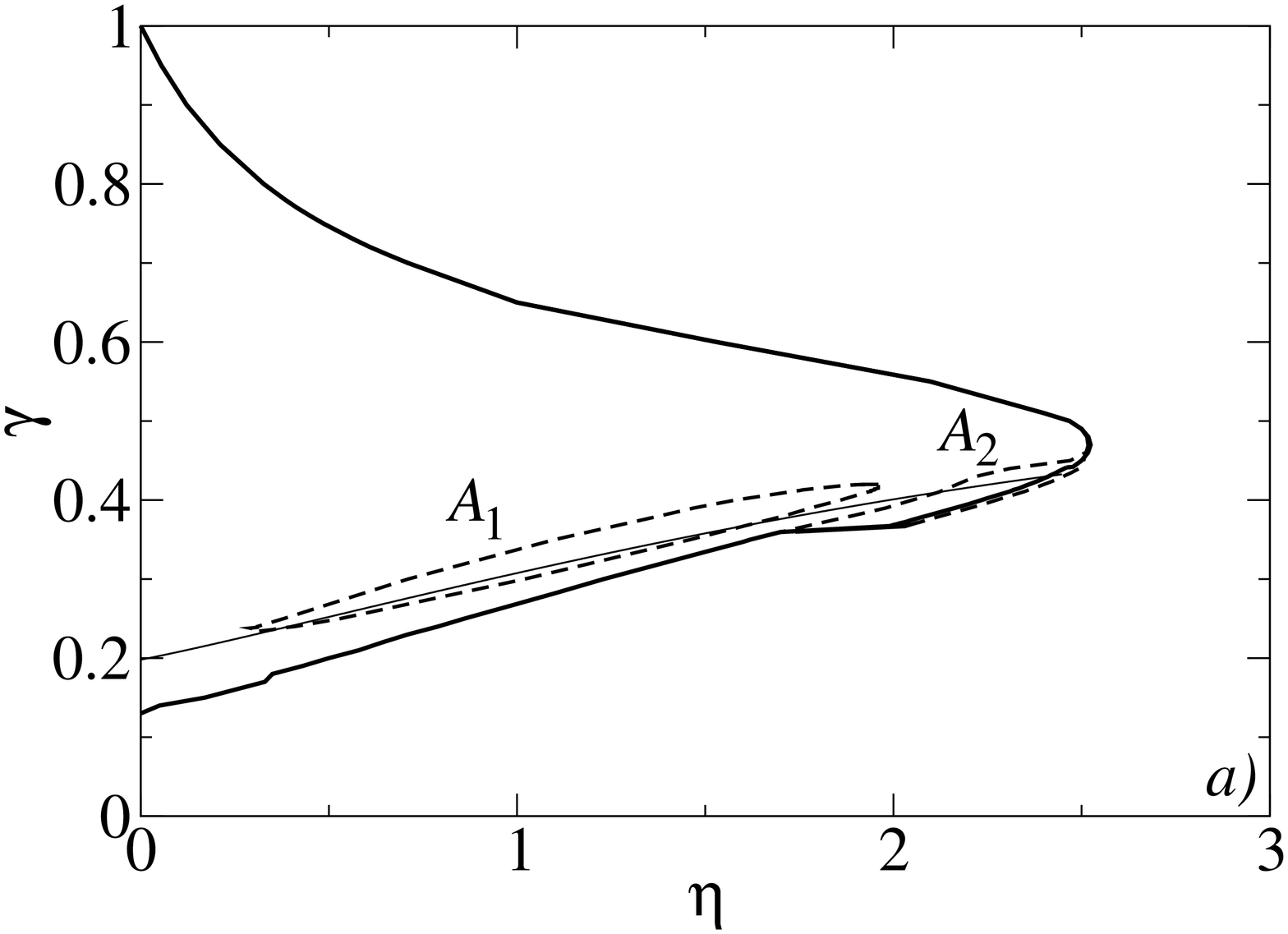,angle=-90,width=9cm,height=8cm}}
\centerline{\psfig{figure=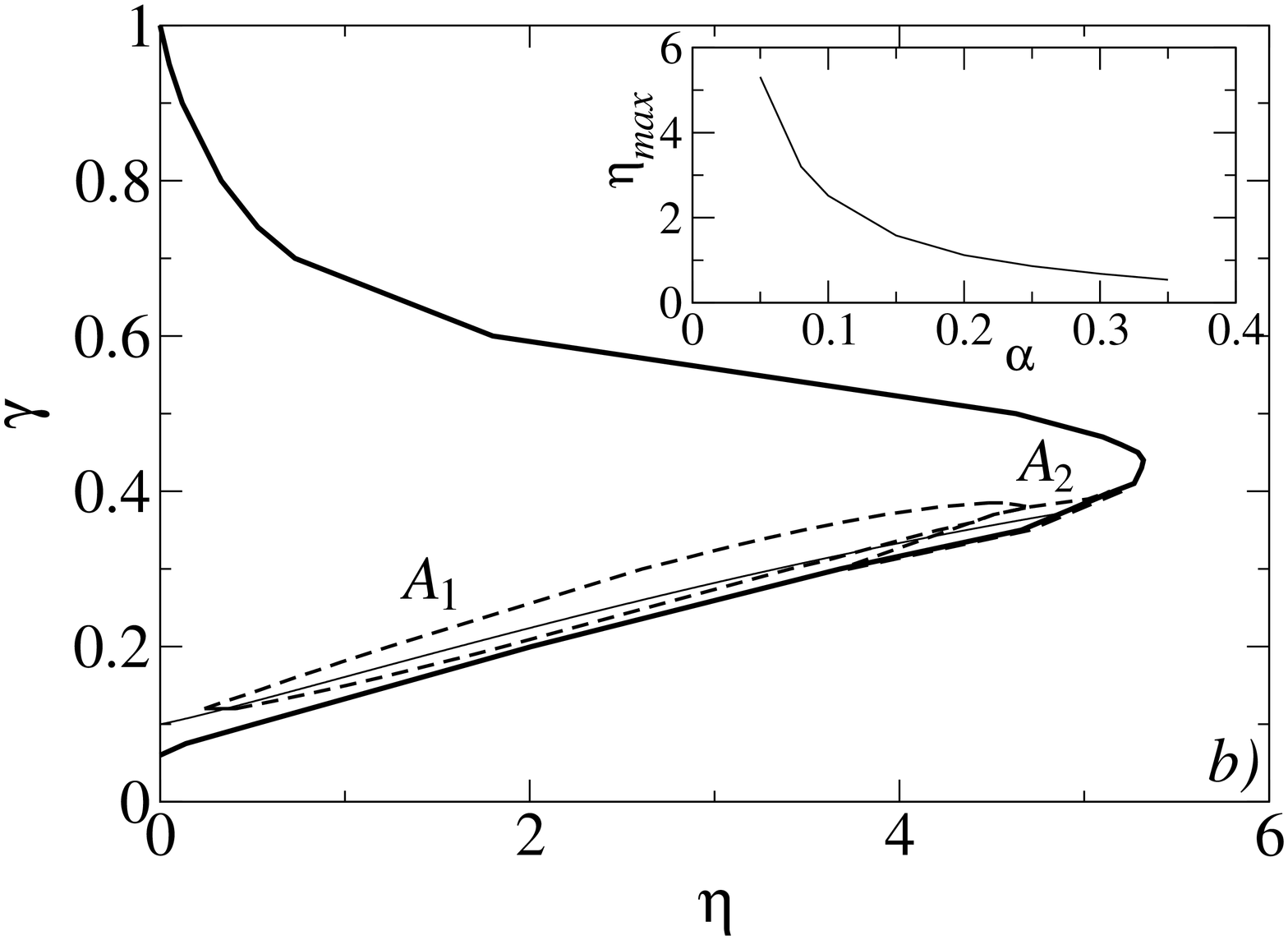,angle=-90,width=9cm,height=8cm}}
\caption{Numerically obtained domains of breather existence 
(thick solid line) and 
instability windows (dashed) lines in parameter space $(\gamma,\eta)$.
The prediction of the breather state instability based on the parametric 
resonance between the breather state and EOs (Eq. (\ref{res2}) for $m=0$) is shown by 
the thin solid line.
The self inductance $\beta_L=0$, and the damping $\alpha$ is 0.1 (Fig. 3a) 
and 0.05 (Fig. 3b). The inset in Fig. 3b shows 
the dependence of the maximum value of anisotropy 
$\eta_{max}$ on the damping $\alpha$. }
\end{figure} 

In order to study the interaction of the breather state with EOs we 
carried out 
direct numerical simulations of the system of equations (\ref{phi1}) and 
(\ref{self}) for different values of anisotropy. Moreover, we monitored the 
current-voltage ($I$-$V$) characteristics  of the system decreasing 
(increasing) the dc bias by a step of $\Delta \gamma~=~0.0005$. \cite{comment2}
The initial values of the dc bias $\gamma$ corresponded approximately to the 
center of the
breather existence domain for the particular value of the anisotropy $\eta$.

We start with the case of a small anisotropy $\eta~=~0.1$ (Fig. 4). The breather 
was excited at $\gamma~=~0.5$ and it persists in a wide range of the
dc bias. The breather state {\it switches} to the HWS 
as $\gamma$ increases beyond some critical value,
and smoothly transits to the superconducting state as 
the dc bias approaches the retrapping current (Eq. (\ref{ir})). 
Note here, that for such small values of anisotropy the dc analysis 
(Eq. (\ref{IV}) and dashed line in Fig. 4) gives a 
good approximation practically for a whole $I$-$V$ curve and it means 
that the resonant interaction of the breather with EOs is very weak.

\begin{figure}[!hbp]
 \centerline{\psfig{figure=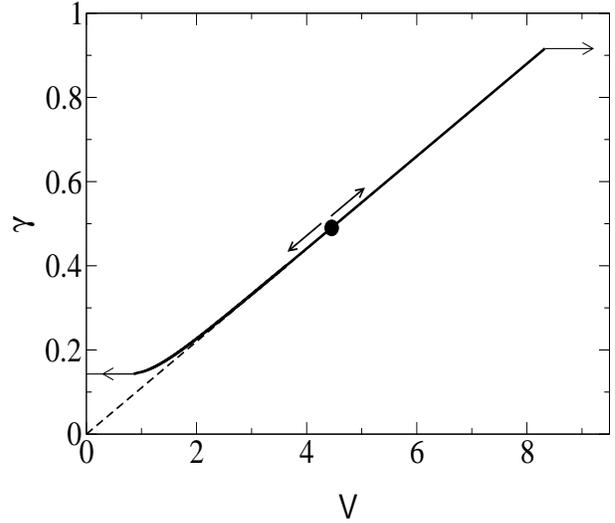,angle=-90,width=9cm,height=8cm}} 
  \caption{$I-V$ characteristics of a breather state for $\beta_L=0$, 
$\alpha=0.1$ and
  $\eta=0.1$. The  $I$-$V$ 
  curve obtained by making use of the dc analysis (dashed line) is 
  also presented. The point indicates 
  the initial value of dc bias and the arrows shows the process of dc bias 
  increase (decrease).
  } 
  \end{figure} 
  
As we increase the value of anisotropy to $\eta~=~1$,  the
$I$-$V$ curve displays peculiar features (Fig. 5). Indeed, as the 
dc bias decreases the $I$-$V$ curve deviates from the linear behaviour 
((Eq. (\ref{IV}) and  dashed line in Fig. 5), displays a cusp  
at $\gamma~=~0.34$
and eventually {\it switches} to the HWS state.  
It is interesting that this scenario is realized if we start from rather 
large values of the dc bias $\gamma~\geq~0.33$. 
For lower dc bias values we observe a second disconnected branch 
(the initial 
value of dc bias was 0.27) which displays a back bending
and switching to the 
HWS in the $I$-$V$ curve as the dc bias was increased. 
With decreasing current this breather state transits to the superconducting 
state.

\begin{figure}[!hbp]
 \centerline{\psfig{figure=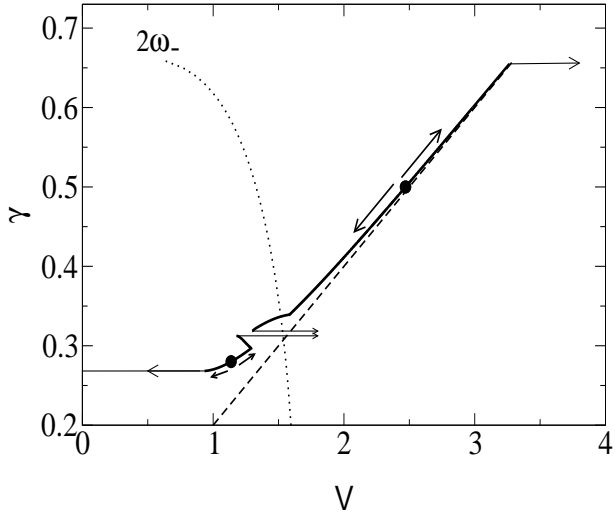,angle=-90,width=9cm,height=8cm}} 
  \caption{$I-V$ characteristics of a breather state for $\beta_L=0$, 
$\alpha=0.1$ and
  $\eta=1$. The dotted line shows the dependence of the 
  characteristic frequency combination 
  of EOs on the dc bias $\gamma$ and the dashed line 
  represent the dc analysis.  
  } 
  \end{figure} 
  
All these features occurring in the current region $0.28~\leq~\gamma~\leq~0.33$,
can be explained as a {\it parametric resonant interaction} between 
the breather state and excited EOs. First, we found a good agreement 
between the breather frequency 
$\Omega~=~\frac{\gamma}{\alpha (1+\eta)}~=~1.6$ and the frequency of 
EOs $2\omega_{-}~=~1.52$ ( for $\gamma=0.32$ ).
To study this resonant interaction more precisely 
we carried out the Floquet analysis of the breather state. 
Fig. 6 illustrates the evolution of the modulus and the argument
of the eigenvalues of the Floquet matrix ${\cal M}$ with $\gamma$.
The arguments of the eigenvalues can be restricted as 
$0~\leq~Arg(\lambda)\leq \pi$. 
As the dc bias decreases we observed that for the particular dc bias
$\gamma=0.34$ an 
eigenvalue of the Floquet matrix crosses the unit circle at $-1$ 
(Fig. 6).
Thus the original breather state becomes {\it unstable} 
(although it continues to exist as a solution of the equations
of motion) and this instability is driven by the 
parametric resonance (Fig. 6b) with the EO of the frequency $\omega_{-}$.
Note, that as the current decreases further this eigenvalue 
returns into the unit circle, the breather state is again stable and 
thus an {\it instability window} of the breather is found (the region $A_1$ in 
Fig.3(a,b)). The position of this instability window in 
the parameter space $\gamma$-$\eta$ can be predicted using Eq. (\ref{res2}) or
in the form (thin solid lines in Fig. 3):
$$
\gamma_m^{inst}(\eta)  = \frac{2\sqrt{2}\alpha^2(1+\eta)^2}{(2m+1)^2} \cdot 
$$
\begin{equation}
\label{ieta}
 \sqrt{ -1 + \sqrt{1 + \frac{(2m+1)^4}{4\alpha^4(1+\eta)^2(1+2\eta)^2}}}
 \;\;\;\;\;\;.
\end{equation}

\begin{figure}[!hbp]
\centerline{\psfig{figure=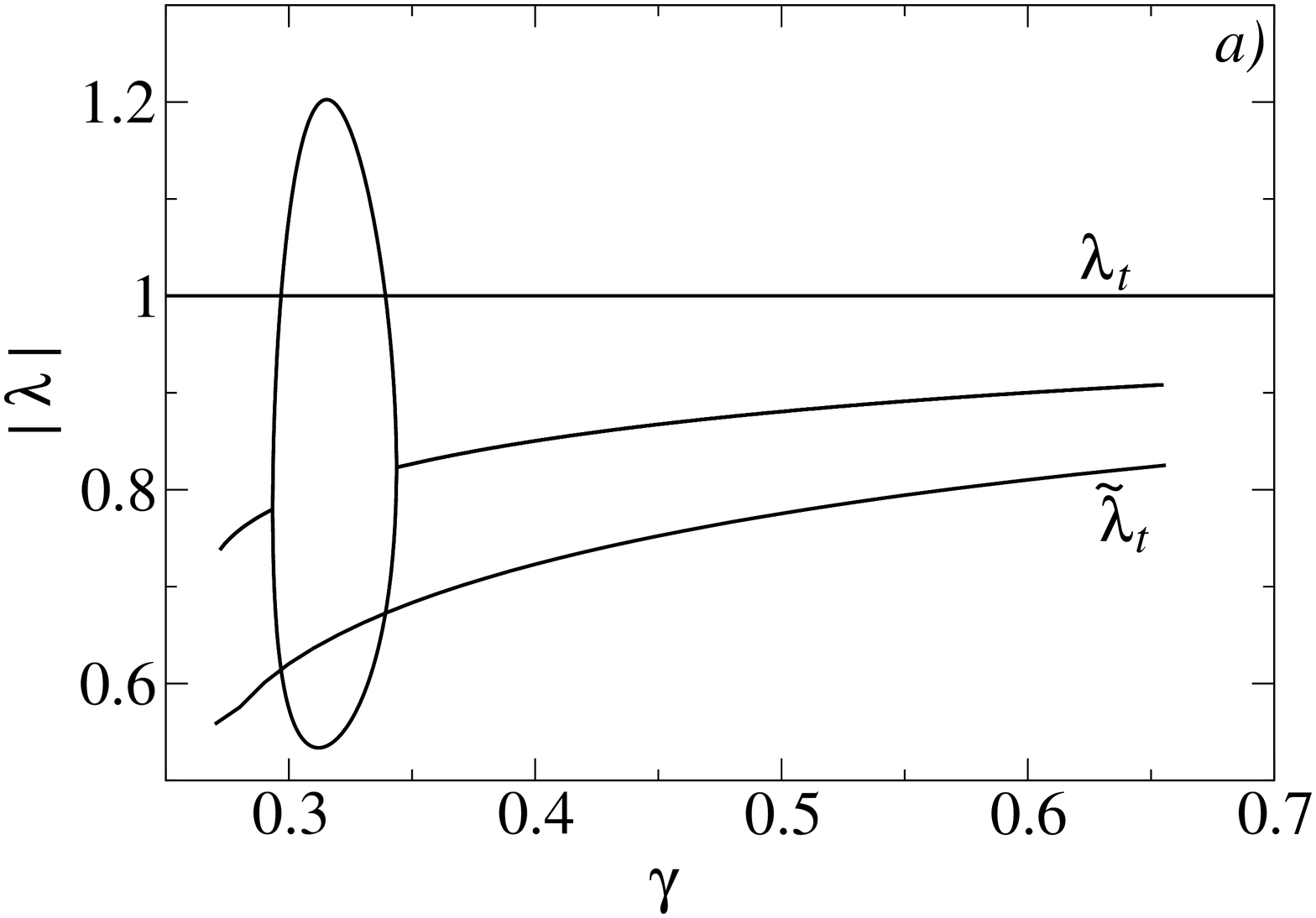,angle=-90,width=9cm,height=8cm}}
\centerline{\psfig{figure=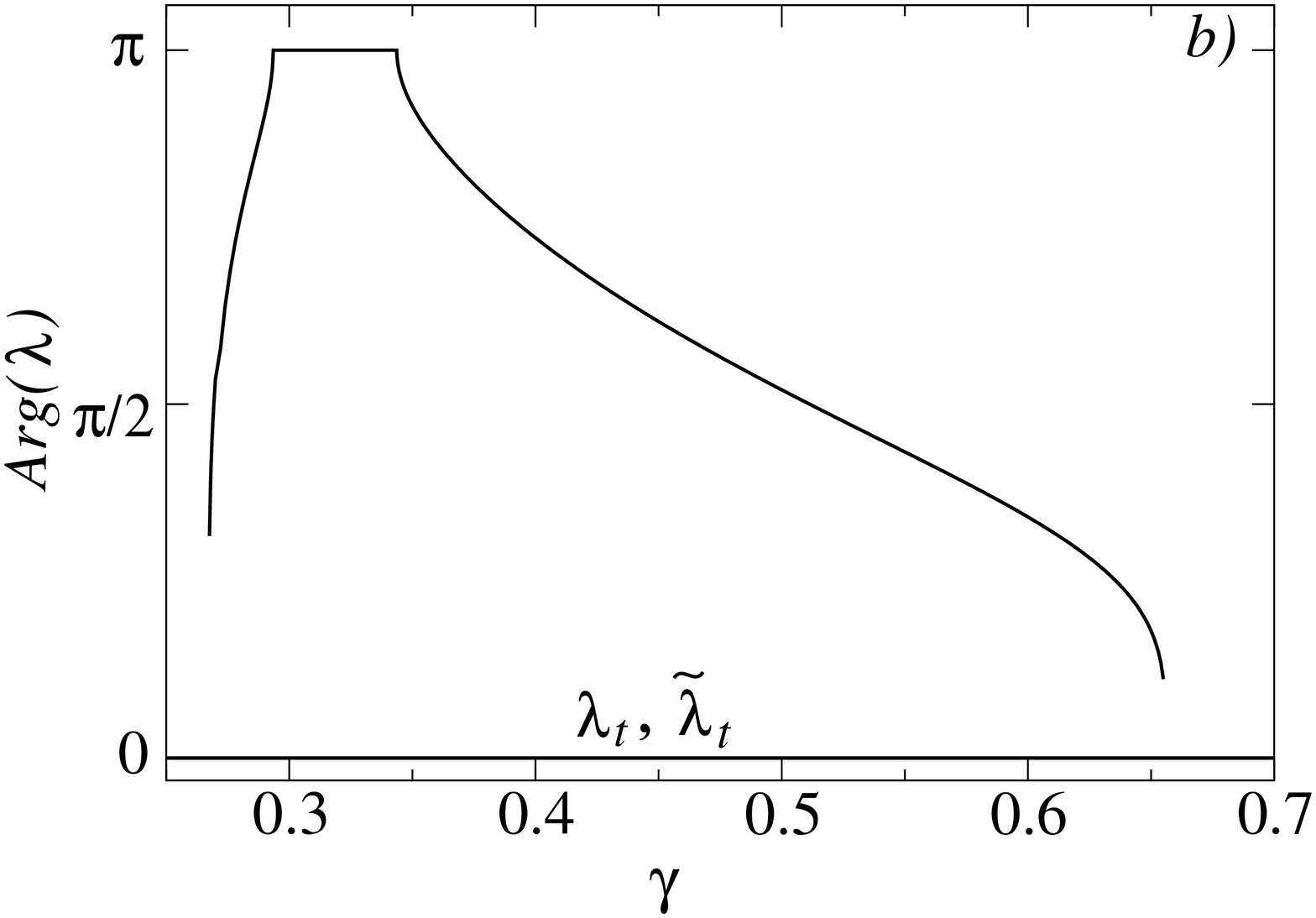,angle=-90,width=9cm,height=8cm}}
\caption{a) Modulus and b) argument of the Floquet multipliers as a 
  function of dc bias $\gamma$. The
parameters are $\beta_L=0$, $\alpha=0.1$ and  $\eta=1$.}
  \end{figure}
  
The appearance of parametric resonance is often accompanied by a
{\it period doubling} of the state mediating the resonance,   
and thus we checked 
the time dependence of the Josephson phase $\phi_1^v (t)$. 
The results for two 
different dc bias values are presented in Fig. 7. 
One can see that for the dc bias $\gamma=0.2965$ 
the period of the Josephson phase 
$\phi_1^v (t)$ is two times less than the period of $\phi_1^v (t)$ 
for the slightly larger value of $\gamma=0.2970$. 
Thus, the "back bending" (and in same manner the"cusp") of
the $I$-$V$ curve (Fig. 5) is related to this period doubling
process.

\begin{figure}[!hbp]
\centerline{\psfig{figure=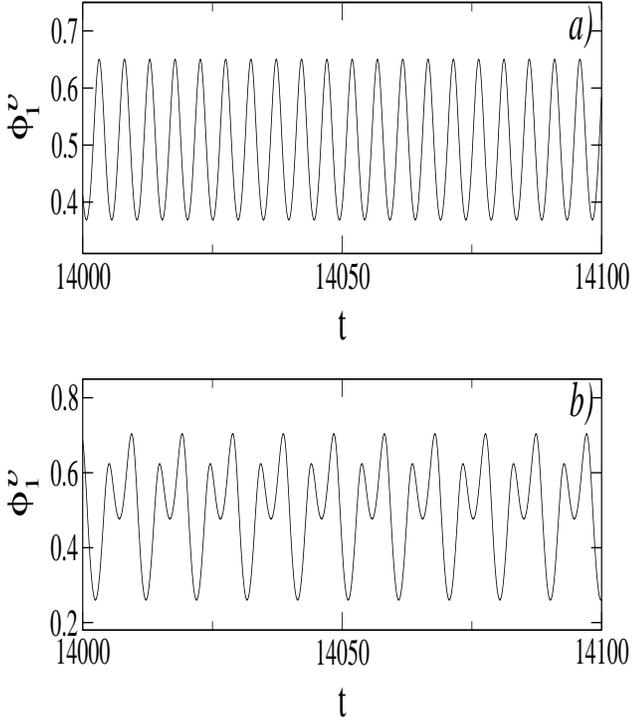,angle=-90,width=9cm,height=12cm}}
\caption{Time dependence of the phase $\phi_1^{v}(t)$ for two different values 
of dc bias $\gamma=0.2965$ (a) and $\gamma=0.2970$ (b). 
The parameters are $\beta_L=0$, $\alpha=0.1$ and $\eta=1$.}
\end{figure} 

As the anisotropy $\eta$ increases and approaches the tip of the 
domain of breather existence (Fig. 3) 
the $I$-$V$ characteristics start to display different features. 
First, for such a large value 
of anisotropy ($\eta~\geq~2$) we observed a large deviation from the 
dc analysis (Fig. 8-10). It indicates 
a strong increase of the amplitudes of EOs. Presumably this
is also the reason why
the switching to the HWS occurs for considerably
smaller $\gamma$ values as compared with the dc analysis 
(see also Ref. [\onlinecite{FistUst}]). 
This circumstance
leads to a shrinking of the domain of breather existence at the tip.

We also found that the switching outcome in this region of anisotropy
upon lowering the dc bias $\gamma$
crucially depends on $\eta$. Indeed, for $\eta~=~2.1$ we observed the 
switching to the superconducting state
(Fig. 8)  but the breather state 
switches to the HWS (Fig. 9)
as the parameter $\eta$ becomes slightly above the value
$2.11$. 
Moreover, for $\eta~=~2.3$ (Fig. 10) 
we observed that the switching to the HWS occurs via a peculiar intermediate 
state. This state is not periodic anymore (either quasiperiodic or chaotic
in time), yet it shows up with typical breather features when averaged over 
time. It was stable over the whole simulation time of $10^6$. 

\begin{figure}[!hbp]
\centerline{\psfig{figure=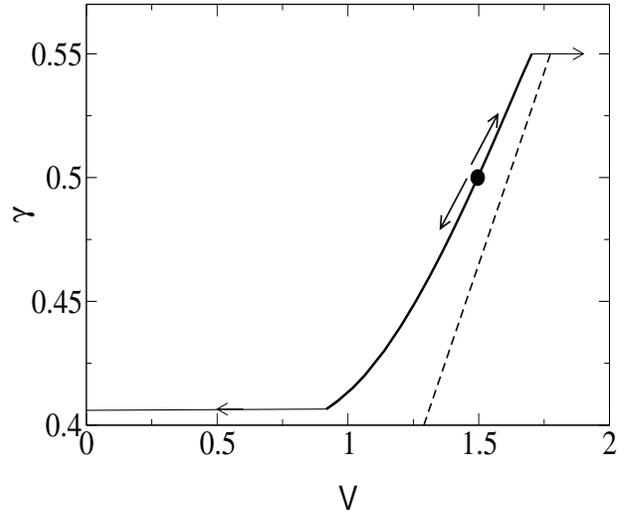,angle=-90,width=9cm,height=8cm}} 
\caption{$I-V$ characteristics of a breather state for $\beta_L=0$, 
  $\alpha=0.1$ and
  $\eta=2.1$. The dashed line represents the dc analysis. 
  } 
  \end{figure} 
  
  \begin{figure}[!hbp]
 \centerline{\psfig{figure=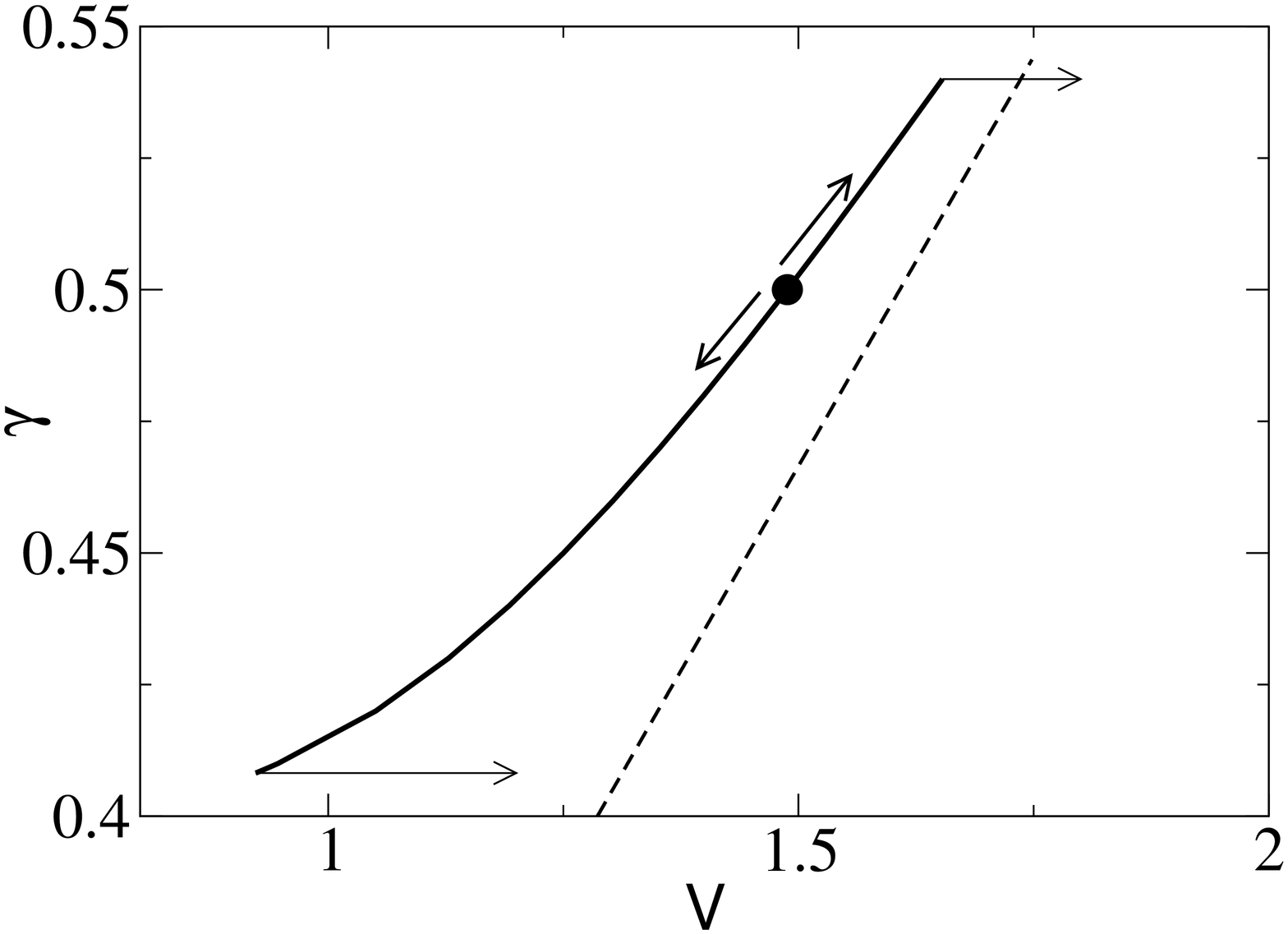,angle=-90,width=9cm,height=8cm}} 
  \caption{$I-V$ characteristics of a breather state for $\beta_L=0$, 
$\alpha=0.1$ and
  $\eta=2.11$. The dashed line represents the dc analysis. 
  } 
  \end{figure}
  
  \begin{figure}[!hbp]
 \centerline{\psfig{figure=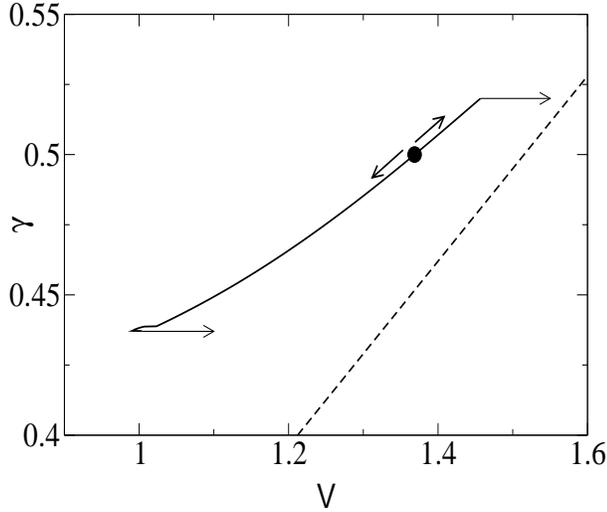,angle=-90,width=9cm,height=8cm}} 
  \caption{$I-V$ characteristics of a breather state for $\beta_L=0$, 
$\alpha=0.1$ and
  $\eta=2.3$. The dashed line represents the dc analysis. 
  } 
  \end{figure}
    
We carried out the Floquet analysis of the breather state and found the 
instability region (region $A_2$ in Fig 3) that is also due to the parametric 
resonance between the breather state and EOs. At variance with the 
considered above instability window $A_1$, this instability region 
adjoins the lower boundary of the breather existence domain.

  \subsubsection {Finite self-inductance case ($\beta_L \neq 0$)}
  The breather properties depend crucially on the self-inductance parameter 
  $\beta_L$. First, we found that the domain of the breather existence shrinks 
  for small and moderate values of $\beta_L$ (see Fig. 11).
  
 \begin{figure}[!hbp]
 \centerline{\psfig{figure=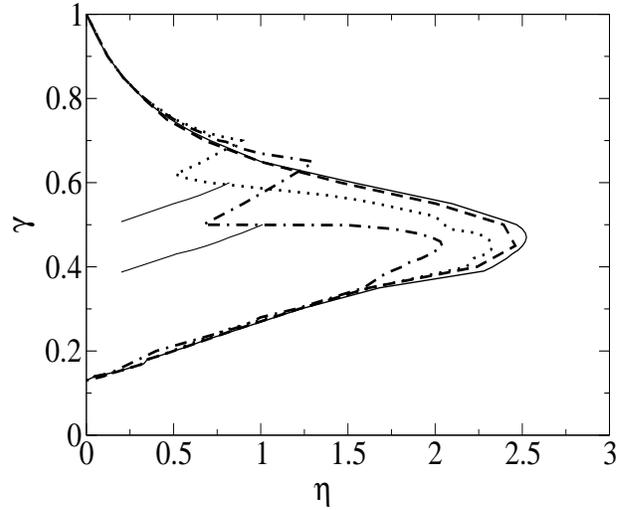,angle=-90,width=9cm,height=8cm}} 
  \caption{Numerically simulated domains of breather existence 
  in parameter space $(\gamma,\eta)$ for: $\beta_L~=~0$ (solid line),
  $\beta_L~=~0.1$ (dashed line), $\beta_L~=~0.3$ (dotted line) and 
  $\beta_L~=~0.5$ (dot-dashed line). The predicted resonances between the 
  breather state and EOs $\Omega=\omega_{+}$ for $\beta=0.3$ and $\beta=0.5$ 
  (from top to bottom) are shown by the thin solid lines. 
  The damping $\alpha$ is 0.1. 
  } 
  \end{figure}
  
  Moreover, the upper boundary of the breather existence domain 
  displays a 
  "saw tooth" form  for moderate values of the anisotropy ($\eta~>~0.5$, 
  see Fig. 11). The reason for such a peculiar shape of the breather 
  existence domain 
  is the resonant interaction between the breather state and EOs 
  (thin solid lines in Fig. 11) leading to large amplitudes of EOs and 
  corresponding ac induced instabilities of the superconducting state of 
  the left vertical junction.
  As the parameter $\beta_L$ increases to large values (see Fig. 12) the 
  resonant interaction between the breather state and EOs becomes weak again. 
In this case the domain of breather existence 
practically coincides with the analytical predictions in Section II 
(see Fig. 2).
   \begin{figure}[!hbp]
\centerline{\psfig{figure=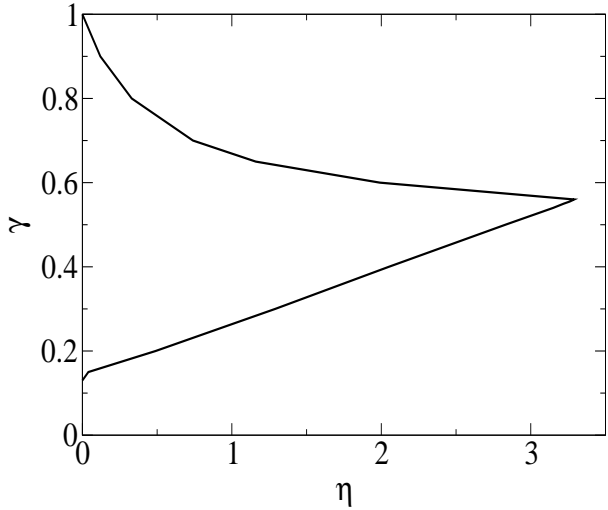,angle=-90,width=9cm,height=8cm}}
\caption{Domain of breather existence in parameter space $(\gamma,\eta)$
  for $\beta_L=100$ and $\alpha=0.1$.}
  \end{figure} 
  
 Also, similar to the case of $\beta_L=0$ we found 
 various instabilities of 
 the breather state driven by the resonant interaction with EOs. These 
 instability windows for a particular case of $\beta_L~=~1$ are shown 
 in Fig. 13. 
\begin{figure}[!hbp]
\centerline{\psfig{figure=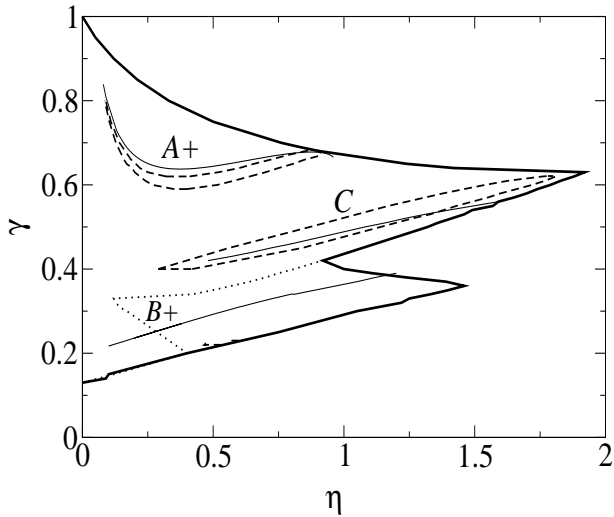,angle=-90,width=9cm,height=8cm}}
\caption{Domain of breather existence and instability windows (dashed line) 
  in parameter space $(\gamma,\eta)$
  for $\beta_L=1$ and $\alpha=0.1$. The predicted resonances between the 
  breather state and EOs are shown by the thin solid lines. The boundary of the 
  primary resonance is shown by the dotted line. }
  \end{figure} 
  The breather instabilities manifest themselves by various features in the 
  $I$-$V$ characteristics. For example, at the value of anisotropy $\eta=0.5$ 
  we observed a rich set of resonant steps and voltage jumps 
  in the $I$-$V$ curve (Fig. 14).
   
  \begin{figure}[!hbp]
\centerline{\psfig{figure=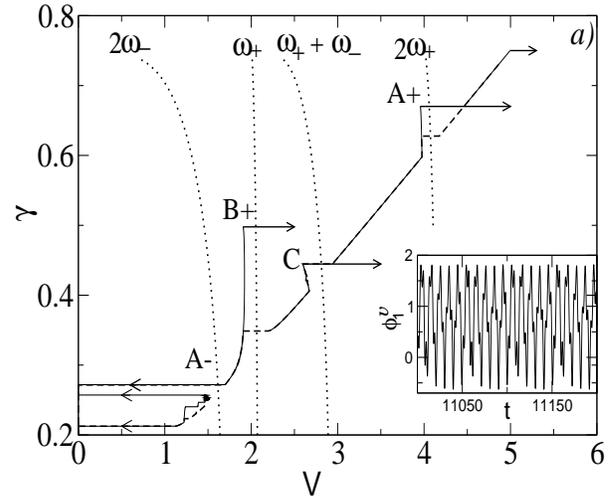,angle=-90,width=9cm,height=8cm}}
\centerline{\psfig{figure=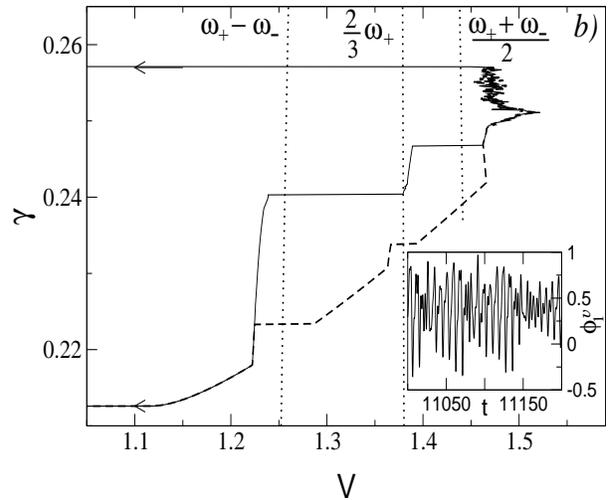,angle=-90,width=9cm,height=8cm}}
  \caption{a) $I$-$V$ characteristics of a breather state for $\beta_L=1$, 
  $\alpha=0.1$ and $\eta=0.5$. Thick dashed lines correspond to
  lowering of the dc bias, thin solid lines to an increasing of the dc bias.
  The labels A-,A+,B+ and C relate the resonances to the observed
  instability windows in Fig. 13.
  b) A zoom of the 
  $I$-$V$ curve. The dependencies of EOs 
  characteristic frequency combinations 
  on the dc bias $\gamma$ are shown by  
  dotted lines. The arrows show the various switching outcomes as the dc bias 
  increases (decreases). The time dependence of the 
  Josephson phase $\phi^{v}_1$ 
  at $\gamma=0.44$, $V=2.61$ and $\gamma=0.25$, $V=1.482$ are shown as 
insets.}
  \end{figure}
  
  For large values of the dc bias ($\gamma~\simeq~0.6$)
  a parametric resonance of the breather state
  with high frequency EOs $\Omega = 2\omega_+$ leads to a 
  resonant step in the $I$-$V$ curve. Moreover, the breather state biased 
  to this resonance displays a period doubling bifurcation 
  which manifests through 
  a sharp corner in the $I$-$V$ curve 
  at the bottom of the resonant step (A+ region).  
  As the dc bias decreases to the value $\gamma~\simeq~0.4$ we observe  
  the combination resonance $\Omega~=\omega_+ +
  \omega_-$.  This resonance is characterized by a large 
  back bending of the $I-V$ curve 
  and the breather state dynamics displays a {\it non-periodic} 
  behaviour (see inset in Fig. 14 and C region).
  
  As the dc bias decreases further to the value of $\gamma~\simeq~0.37$,
  a large resonant step in the 
  $I$-$V$ curve appears. It corresponds to the primary resonance with 
   $\Omega = \omega_+$. 
  This resonant breather state switches 
  to the superconducting state at the value of $\gamma~=~0.27$ as the result of 
  an instability driven by the parametric resonance with low 
  frequency EOs $\Omega = 2\omega_-$. 
  These features are clearly seen in the dependence of 
  Floquet multipliers on the dc bias (see Fig. 15).
  \begin{figure}[!hbp]
\centerline{\psfig{figure=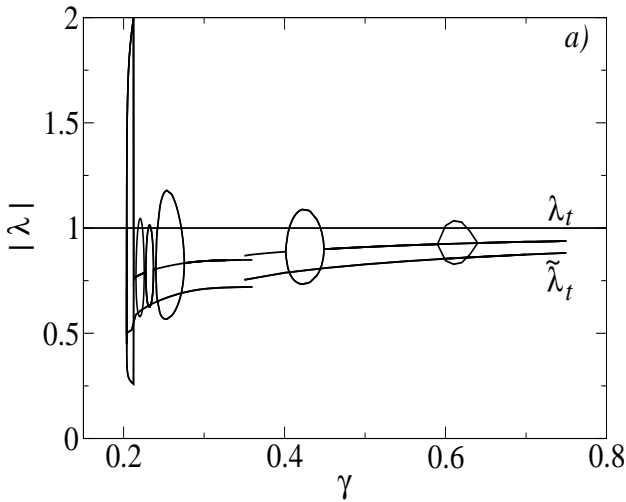,angle=-90,width=9cm,height=8cm}}
\centerline{\psfig{figure=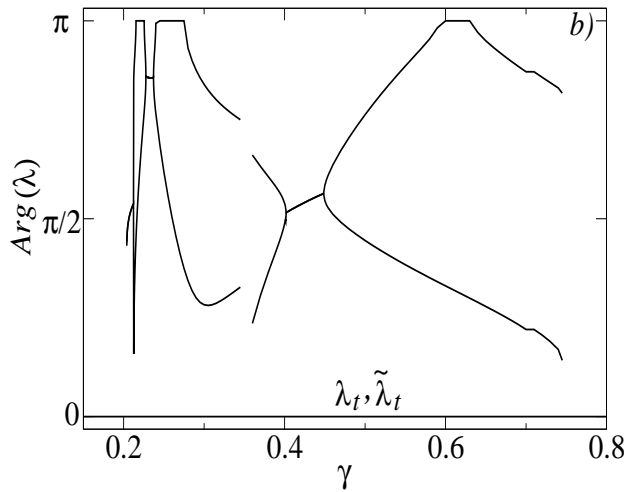,angle=-90,width=9cm,height=8cm}}
  \caption{a) Modulus and b) argument of the Floquet multipliers as a function 
of
  dc bias  $\gamma$. The parameters are $\beta_L=1$ $\alpha=0.1$ and 
$\eta=0.5$.}
  \end{figure}
  The type of resonance can be easily detected from the behaviour
  of the arguments of the Floquet multipliers. While unstable solutions
  (with absolute values of multipliers being larger than one) 
  which corresponding to parametric resonances are characterized
  by 'frozen' arguments at value $\pi$, combination resonances
  are characterized by a merging of arguments over a particular
  interval of dc bias.
    
  For even smaller values of the dc bias, namely
  $0.21~<~\gamma~<0.26$) the breather state becomes again stable. 
  For $\gamma~\simeq~0.26$ this breather state 
  shows up with a non-periodic behaviour (see 
  inset in Fig. 14 (b)). Most important we observe
  three new resonances: a combination resonance
  $\Omega=\omega_+ - \omega_-$, a parametric resonance
  $\Omega = \frac{2}{3}\omega_+$ and yet another combination
  resonance $2\Omega = \omega_+ + \omega_-$. 
  The breather finally switches  to the 
  superconducting state at 
  $\gamma~=~0.21$. 
  
\section{Conclusion}
We have presented an analytical and numerical study of inhomogeneous 
dynamic states (breathers) in a single plaquette consisting of three small 
underdamped Josephson junctions. We found that as the damping $\alpha$ is small
the breather state as a solution of dynamic equations exists 
in a large region of parameters: dc bias $\gamma$ and anisotropy $\eta $
(Fig. 2, 3, 12 and 13). For small and moderate values of 
anisotropy $\eta$ ($\eta~\leq~1$) the dc analysis that completely 
neglects the breather induced EOs (Eqs. (\ref{phiapprox})-(\ref{hws})), 
results in a good estimation of the boundary of breather existence. However, 
we found that the domain of the breather existence shrinks as the anisotropy 
$\eta$ increases, and moreover, for moderate values of the self inductance 
$\beta_L$ the upper part of the boundary displays a "saw tooth" feature.
These effects are due to the excitation of EOs with large amplitudes.

The second major result of the presence of 
EOs is that the domain of the breather existence contains a
large number 
of {\it instability regions}. The physical origin of these instabilities
is the resonant interaction 
between the breather state and the EOs. It is important that due to the 
presence of two different characteristic frequencies of EOs (for the dependence 
of $\omega_{\pm}$ on the self inductance and anisotropy, see 
Eqs. (\ref{omega34})), a rich set 
of resonances (primary, parametric or combination ones) occurs. 
Such resonant interactions can completely destroy the breather state and 
correspondingly lead to significant shifts of the 
switchings of the breather state 
to the HWS or the superconducting state. These switchings appear through 
voltage jumps in the $I$-$V$ curves. Note here, that in most cases 
we observed the switching to the superconducting state as a result of 
breather instability (parametric or combination) but not due to the 
standard retrapping mechanism (Eq. (\ref{ir})). Similar findings have
been also reported 
for 
large Josephson junction ladders. \cite{MFFZP}
 
The most interesting 
observation is that the breather state can survive the resonant instabilities 
by increasing its dynamic complexity. This dynamic complexity manifests 
itself by resonant steps of various types in the $I$-$V$ curves. 
Indeed, we observed a {\it 
resonant } type breather as a result of a primary resonance with 
EOs ($B_+$ region in Fig. 14a), a period doubling process resulting from 
the parametric resonant interaction 
(Fig. 5, regions $A_+$, $A_-$ in Fig. 14) and even a non-periodic 
(quasi periodic or chaotic) behaviour in parameter regions where 
combination resonances appear ($C$ region and the region of small current 
in Fig. 14).

Finally, all observed effects (switching outcomes, instabilities)
are very sensitive to the damping parameter $\alpha$, and different 
combinations of the resonances can appear as the damping parameter $\alpha$ 
decreases. 

What can we learn about breather states in extended ladder geometries?
Our analysis shows that stable breathers can be induced for
large self inductance. If on the other hand
one needs switchings between different
breather states and resonances, intermediate $\beta_L$ values should be chosen.
In addition variations of the damping (e.g. via temperature control)
may increase the complexity of expected switching scenarios.
At the same time it should be clear that if already such a small system
(with three degrees of freedom) shows up with a high complexity of
the phase space structure, the situation for large numbers of degrees
of freedom as for Josephson ladders will be even more complex.
\\
Acknowledgements. 
We thank J. J. Mazo, A. Miroshnichenko, J. Page, M. Schuster, A. Ustinov
and Y. Zolotaryuk for useful discussions. This work was supported by the 
European Union under the RTN project LOCNET HPRN-CT-1999-00163.


\begin{thebibliography}{9}
\bibitem{comment2} As it was observed in Ref. [\onlinecite{MFFZP}] 
   for the Josephson ladder case, tiny steps $\Delta \gamma$ and large 
   integration times have to be used in order to observe various 
   switchings in the $I$-$V$ curves.
\end{thebibliography}
  \end{document}